\documentstyle[12pt,epsfig]{article}

\begin{document}

{\noindent \small Physica A {\bf 278}, 579 (2000).}
\begin{center}
{\LARGE \bf Strategy Selection in the Minority Game}

\vskip 0.6cm
{\large R. D'Hulst \footnote{\noindent corresponding author: Rene.DHulst@brunel.ac.uk\\ phone: +44-(0)1895-274000; fax: +44-(0)1895-203303} and G.J. Rodgers}
\vskip 0.4cm

\it Department of Mathematical Sciences, Brunel University,\\
Uxbridge, Middlesex, UB8 3PH, UK.
\vskip 0.5cm

\end{center}
\setlength{\unitlength}{1cm}

{\noindent \large Abstract}
\vskip 0.4cm

We investigate the dynamics of the choice of an active strategy in the minority game. A history distribution is introduced as an analytical tool to study the asymmetry between the two choices offered to the agents. Its properties are studied numerically. It allows us to show that the departure from uniformity in the initial attribution of strategies to the agents is important even in the efficient market. Also, an approximate expression for the variance of the number of agents at one side in the efficient phase is proposed. All the analytical propositions are supported by numerical simulations of the system.


\newpage
\section{Introduction}
\label{sec:introduction}

The minority game, introduced by Challet and Zhang \cite{challet9798}, is a simple model to mimic the internal dynamics of a market where agents can only buy or sell one commodity. At each time step, $N$ agents choose independently between two different sides, 0 or 1, that is, they have to choose recurrently between two different opportunities, like buying or selling. The side chosen by the minority of agents is identified as the winning side and all agents that have chosen this side are rewarded with a point. The points of the other agents, the losers, do not change. Agents strive to maximize their number of points. Such minority competitions are reminiscent of systems where individuals have to compete for scarce resources. Ref. \cite{zhang98} motivates this model in more detail. Various extensions of the model have also been introduced \cite{tous}.

The record of which side was the winning side for the last $m$ time steps constitutes the history of the system. Agents analyze the history for the last $m$ time steps to make a decision for the next time step. Hence, they have to be provided with a vector of length $2^m$ whose components are 0's or 1's. Each component is a decision corresponding to a particular history. Any such binary vector of size $2^m$ is called a strategy. See \cite{zhang98} and references herein for a more detailed discussion about the strategies used by the agents.

Agents use a strategy at each time step. Each agent has at his disposal a fixed set of $s$ strategies chosen at random, multiple choices of the same strategy being allowed. At each time step, an agent uses one of his strategies to make a decision. The chosen strategy is called the active strategy while, conversely, the $s-1$ other strategies are called passive strategies.

The strategies, like the agents, are rewarded points according to their merit. A strategy earns a point each time it forecasts the winning side, irrespective of whether it is used or not. The points given to a strategy are called virtual points, they represent the points an agent could have earned had he always played with this strategy. Consequently, an agent always plays with his strategy with the highest number of virtual points, assuming that a strategy that has already won will win again. This assumption neglects the influence of the agent himself on the result.
 
It has been shown numerically that the system exhibits a transition from an efficient to an inefficient phase \cite{savit97}, the transition occuring when the number of different histories $2^m$ is of the order of the number of agents $N$. The parameter $\alpha = 2^m/N$ has been introduced to describe universality in the model. The transition from an efficient to an inefficient market takes place at a critial value $\alpha_c$. In the efficient phase, any information is promptly used by all the agents while arbitrage opportunities exist in the inefficient phase. That is, the virtual points given to the strategies contain information in the inefficient phase, which allows the agents to perform better on average than a system with agents guessing at random between the two sides. On the contrary, the agents are winning less points on average in the efficient phase.

Although the model is quite simple, no exact analytical solution has been proposed, the main difficulty originating from the dynamical process through which agents choose their active strategy. In fact, a stochastic model with agents choosing their strategy at random among their set of $s$ strategies has also been introduced \cite{dhulst99}. In this simplified case, all the dynamics of the model is included in the geometry of the space of strategies, and analytical results using the Hamming distance between strategies are accessible. The solution of a closely related model has been proposed \cite{challet99-2}. This latter model coincides with the minority game in the inefficient phase and also exhibits a transition at $\alpha_c = 0.33740...$ The aim of this paper is to address the dynamics of the choice of a strategy by the agents.

\section{The history distribution}
\label{sec:history distribution}

Consider a particular history $h$. Suppose that after $t$ time steps, this history has happened $n$ times. Among these $n$ occurrences, side 0 has won for $n_0$ time steps and side 1 has won for $n_1$ time steps, with $n_0 + n_1 = n$. Let us define $x_h = | n_0 - n_1 |$ as a measure of the asymmetry between the two sides for the history $h$. If $n_0=n_1$ for a history $h$, that is, $x_h =0$, each side has won the same number of times. The virtual points of the strategies do not contain any information about this history and the two sides are indistinguishable. On the contrary, for $x_h \not = 0$, one side has won more often than the other and the symmetry between the two sides is broken. When $x_h \not = 0$, the agents try to play with the strategies that forecast the side that has won more often for $h$. This behaviour is restricted by the number $s$ of strategies available to an agent and by the size of the memory $m$, because the virtual points contain information about all the histories. When $m$ increases, the number of histories increases so that the agents are less and less able to retrieve information for a particular history.  

We define the history distribution $D (x )$ as the probability that a history chosen at random is associated to $x$. $2^m D (0)$ is the average number of histories with $x = 0$. For these histories, the two sides are symmetric. If such a history happens, the system behaves like one where the agents choose between the two sides at random.

Fig. 1 (a) and (b) present numerical results for $D (x)$ as a function of $\alpha$ for agents having $s=2$ strategies at disposal. For $\alpha < \alpha_c$, most of the distribution $D (x)$ is concentrated around the two values $x = 0$ and $x = 1$. For these values of $\alpha $, when a history $h$ happens, agents guess at random on alternate time steps. On the other time steps when the history $h$ happens, they try to go to the side that won the last time the same history happened \cite{savit97}. In this sense, the behaviour of the system is antipersistent \cite{challet99-2}. Around $\alpha = \alpha_c$, the agents are less and less able to efficiently retrieve information from the virtual points. For some histories the asymmetry between the two sides is large. In fact, for $\alpha > \alpha_c$, we observe numerically that the history distribution is no longer stationary.  Fig. 1 (b) presents numerical results for the average value of $D (x)$ after 10000 time steps. If the simulations are continued for longer times, $D (x)$ tends towards a flat distribution, every value of $x$ being equally likely to occur.

Assuming a continuous time evolution as well as a continuous variation of $D (x)$ over non integer values of $x$, the history distribution $D (x)$ for $x\ge 2$ evolves as

\begin{equation}
{\partial D (x,t)\over \partial t} =  {\partial \over \partial x}((1-2p^{+} (x, t)) D(x, t)) + {1\over 2}{\partial^2 D (x,t)\over \partial x^2}.
\end{equation}
where $p^{+} (x)$ is the probability that if a history associated with $x$ is selected, $x$ will increase by one. The history distribution can reach a stationary state only if $p^{+} < 0.5$. The first term on the right hand side of the previous equation describes the local balance for the history distribution while the second term on the right hand side is equivalent to a surface tension. It smoothes the curve, but can be neglected if the distribution does not reach a stationary state. The equation for $x = 0$ is 

\begin{equation}
2^m {\partial D (0) \over \partial t}Ê= - D (0) + (1- p^{+}(1)) D (1).
\end{equation}
and the equation for $x = 1$ is 

\begin{equation}
2^m {\partial D (1) \over \partial t}Ê= D (0) - D (1) + (1- p^{+}(2)) D (2).
\end{equation}
For $p^{+} (1) =0$, $D(x\ge 2) =0$ and $D(0)=D(1)=1/2$, as observed for the smallest values of $\alpha$.

Fig. 2 (a) and (b) present numerical results for $p^{+}$ as a function $x$ for various values of $\alpha$, with 2 strategies per agent. For $\alpha \ll \alpha_c$, the system is particularly antipersistent, with $p^{+}$ going to 0 as $x$ is increased. For $0 < \alpha < \alpha_c$, $p^{+} < 0.5$ for the smallest values of $x$, typically for $x$ ranging from 1 to 10. When $x$ is increased, $p^{+} \rightarrow 1$, with huge fluctuations as reported on Fig. 2 (a) for $\alpha = 0.3$. As can be seen from Fig. 1 (a), just a few histories are associated with these high values of $x$. Hence, for $0 < \alpha < \alpha_c$, the system is mainly antipersistent with some exceptions. For these particular histories, a large number of agents are not able to adapt, because both their strategies make the same prediction. Consequently, the symmetry between the two sides is broken. For $\alpha > \alpha_c$, most histories are persistent, with $p^{+} > 0.5$. In this case, the history distribution is no longer stationary, as mentioned above.

To test the stationarity of the distribution in the efficient phase, we present at Fig. 3 the relative difference between the number of histories $N^{-} (x)$ going from $x$ to $x-1$ and the number of histories $N^{+} (x-1)$ going from $x-1$ to $x$. $b (x)$ is defined as

\begin{equation}
b (x) = {N^{-} (x) - N^{+} (x-1)\over N^{-} (x) + N^{+} (x-1)}.
\end{equation}
Numerical results for $b(x)$ are presented at Fig. 3 for $\alpha = 0.3$ and $s=2$. For the small values of $x$, $b (x) \simeq 0$, showing that the distribution has reached a stationary state. On the contrary, for larger values of $x$, $b (x) \rightarrow -1$. This shows that for some histories, the symmetry is already broken before the transition. The increasing number of these kinds of histories as $\alpha$ is increased towards $\alpha_c$ will destabilize the history distribution, finally becoming non-stationary at $\alpha_c$. Already for $\alpha < \alpha_c$, the history distribution is not stationary for the higher values of $x$. 

\section{Before the transition}
\label{sec:before the transition}

In this section, we restrict our attention to the low values of $\alpha$, that is, to the values of $\alpha$ such as $p^{+} (1)$ is approximately 0. This range of values for $\alpha$ will be refered as the low $\alpha$ phase, typically, $\alpha < 0.1$. In this case, only histories with $x =0$ or 1 have to be considered. If there are $k$ histories with $x = 1$, the strategies can be classified in decreasing order of virtual points \cite{johnson98-1}, giving $k+1$ different ranks. Strategies of rank 0 have the highest number of virtual points, while strategies of rank $r$ have $r$ less points, with $r\in \lbrack 0,k \rbrack$. There are

\begin{equation}
K_r (k) = {2^{2^m} k!\over 2^k r! (k-r)!}
\end{equation}
strategies on rank $r$. $p_r (k) = K_r (k)/2^{2^m}$ is the probability to choose a strategy of rank $r$ when picking a strategy at random in the strategy space. 

Comparing two strategies of rank 0, these strategies have always made the same prediction for any history with $x =1$. Comparing a strategy of rank 0 with a strategy of rank $r$, these strategies made the same predictions for all the histories with $x =1$, except for $r$ such histories. A reduced Hamming distance $d^{*}$ between two strategies $\sigma$ and $\sigma'$ can be defined as 

\begin{equation}
d^{*} = {1\over k} \sum_{k\in V} | \sigma_h - \sigma'_h |
\end{equation}
where $V$ is defined as the set of histories $h$ with $x \not = 0$. Then, any strategy at rank $r$ is at a reduced distance $d^{*} (r) = r/k$ from any strategy of rank 0. This means that if a history with $x = 1$ happens, $d^{*} (r)$ is the probability that a strategy of rank $r$ and a strategy of rank 0 will make opposite predictions.

As agents always try to use the strategy with the highest number of virtual points they have, there should be on average $N_r^{(0)}$ agents in group $r>0$, with \cite{dhulst99,johnson98-1}

\begin{equation}
{N_r^{(0)}\over N} = \left( 1 - \sum_{j=0}^{r-1} p_j (k)\right)^s - \left( 1 - \sum_{j=0}^{r} p_j (k)\right)^s
\label{eq:classification 0}
\end{equation}
with $N_0^{(0)}/N = 1 - (1-2^{-k})^s$. The superscript $(0)$ refers to the first approximation used for the classification of agents. In fact, the distribution of strategies according to their virtual points is somehow the result of the departures from uniformity present in the initial choice of strategies by the agents. Eq. (\ref{eq:classification 0}) neglects this property.

To go beyond the $N_r^{(0)}$ approximation, consider the average number of winners $N_w$ for agents playing at random. This is equal to 

\begin{equation}
N_w = {N\over 2} - N {(N-1)!\over 2^N \left( {N-1\over 2}\right)!^2}.
\end{equation}
In this case, the probability to win per agent is $\tau_w = N_w / N$. If there are $k$ histories with $\Delta n =1$, there are on average

\begin{equation}
Q_r = N {k!\over r! (k-r)!} \tau^{k-r} (1-\tau)^{r}
\end{equation}
agents that possess a strategy of rank $r$, if we suppose that the agents have never adapted. Then, we let the agents adapt if they have among their $s-1$ remaining strategies a strategy of a lower rank. The next approximation for the average number of agents $N_r^{(1)}$ on rank $r$ is given by

\begin{eqnarray}
\nonumber 
N_r^{(1)} &=& Q_r \left( 1 - \sum_{j=0}^{r-1} p_j \right)^{s-1} + \left( N - \sum_{j=0}^{r} Q_r \right) \\
&\times& \left(\left( 1 - \sum_{j=0}^{r-1} p_j (k) \right)^{s-1} - \left( 1 - \sum_{j=0}^{r} p_j (k)\right)^{s-1}\right).
\label{eq:classification 1}
\end{eqnarray}
Note that $N_0^{(1)} = Q_0 + (N - Q_0) (1 - (1-2^{-k})^{(s-1)})$. In this approximation, we suppose that the distribution of the strategies reflects the property of the active strategy of an agent, but does not contain any information about the $s-1$ passive strategies.

With the notation introduced in this section, the variance of the number of agents at one side, $\sigma^2$, in the low $m$ phase is given by

\begin{eqnarray}
\nonumber
\sigma^2 &=& \sum_{k=0}^{2^m} {(2^m)!\over k! (2^m-k)!} D (0)^{k} (1- D(0))^{2^m-k} \left( {k N\over 2^{m+2}} + \left({2^m-k\over 2^m}\right)\right. \\
&\times& \left. \left((\sum_{r=0}^{k} N_r d^{*}_r - N/2)^2 + \sum_{r=0}^{k} N_r d^{*}_r (1 - d^{*}_r) \right)\right).
\label{eq:sig2-low}
\end{eqnarray} 
In this expression, it is considered that every history is equally likely to occur. The frequency of occurence of a history with $x =0$ is equal to $D(0)$. For every such history, agents are guessing at random, giving a contribution to $\sigma^2$ of $N/4$. Histories with $x =1$ occurs with a frequency equal to $D(1) = 1-D(0)$. For every such histories, agents are using the lowest ranked strategy they possess. Agents at rank $r$ go to the side that lost more often for the history considered with a probability $d^{*}_r$, so that on average $N_r d^{*}_r$ agents are going to this side. The variance of this number is equal to $N_r d^{*}_r (1-d^{*}_r)$. 

The result of Eq. (\ref{eq:sig2-low}) as a function of $\alpha$ is presented in Fig. 4 using $N_r^{(0)}$ and $N_r^{(1)}$, for agents having $s=2$ strategies at their disposal. We have checked that we obtain similar results for a given value of $\alpha$, using several values for $N$ and $m$. For a fixed value of $N$, Eq. (\ref{eq:sig2-low}) is a very slowly increasing function of $m$. The variance $\sigma^2$ of the number of agents at one side obtained by numerical simulations are also reported for comparison, using $m=4$. Numerically, $\sigma^2 / N$ is also slightly increasing with $m$ for a fixed value of $N$. As can be seen, Eq. (\ref{eq:classification 1}) gives a better agreement with the numerical simulations, up to $\alpha = 0.1$. 
   
\section{Conclusions}
\label{sec:conclusions}

In this paper, we introduced the history distribution as a complementary tool to the Hamming distance to investigate the dynamics of the minority game. We argue that the history distribution is useful to investigate the process of the choice of the active strategy of an agent. The properties of the history distribution are numerically investigated, showing that when the transition from an efficient to an inefficient market occurs, the distribution becomes non-stationary. 

An approximate analytical expression for the variance of the number of agents at one side is proposed for the efficient phase. It is only exact for very low values of $\alpha$, typically, $\alpha < 0.1$. We were able to show that the agents are not just picking their strategies at random in a ranking of the strategies according to their virtual points. In fact, the ranking is the result itself of the initial choice of the strategies by the agents. Hence, the information about the strategies possessed by the agents has to be retrieved from the ranking.

This work can be seen as complementary to the work of Challet and Marsili \cite{challet99-2}. It supplies some analytical considerations in the efficient phase, where an analytical solution is still lacking.

\newpage
{\noindent \large Figure Captions}

\vskip 1.0cm
{\noindent \bf Figure 1} --- Numerical simulations for the history distribution $D (x)$ as a function of $\alpha$ for agents with $s=2$ strategies at disposal. (a) $\alpha_c > \alpha=0.1$, 0.2 and 0.3; (b) $\alpha_c < \alpha = 0.4$, 0.5 and 0.6.

\vskip 1.0cm
{\noindent \bf Figure 2} --- Numerical simulations for the parameter $p^{+}$ as function of $x$ for agents with $s=2$ strategies at disposal. (a) $p^{+}$ for $\alpha = 0.3$; (b) $p^{+}$ for small values of $x$ and $\alpha = 0.1$, 0.2, 0.3, 0.4, 0.5 and 0.6. Note the progressive disappearance of the dip in the curve as $\alpha$ is increased.

\vskip 1.0cm
{\noindent \bf Figure 3} --- Numerical simulations for the stability of the history distribution for $\alpha = 0.3$, $s=2$. $b$ is proposed as a function of $x$. The curve is flat for small values of $x$, but as can be seen in the inset, $b \rightarrow -1$ as $x$ is increased. 

\vskip 1.0cm
{\noindent \bf Figure 4} --- Variance $\sigma^2$ of the number of agents at one side divided by the number of agents $N$, as a function of $\alpha$ for $s=2$. Numerical results  ($\bullet$) from simulations of the system are compared to the analytical results. The analytical results for $\sigma^2/N$ using the distributions $N_R^{(0)}$ ($\circ$) and $N_R^{(1)}$ ($\triangle$) are presented.
\newpage

\begin{picture}(21,27)(3,-4)
\epsfig{file=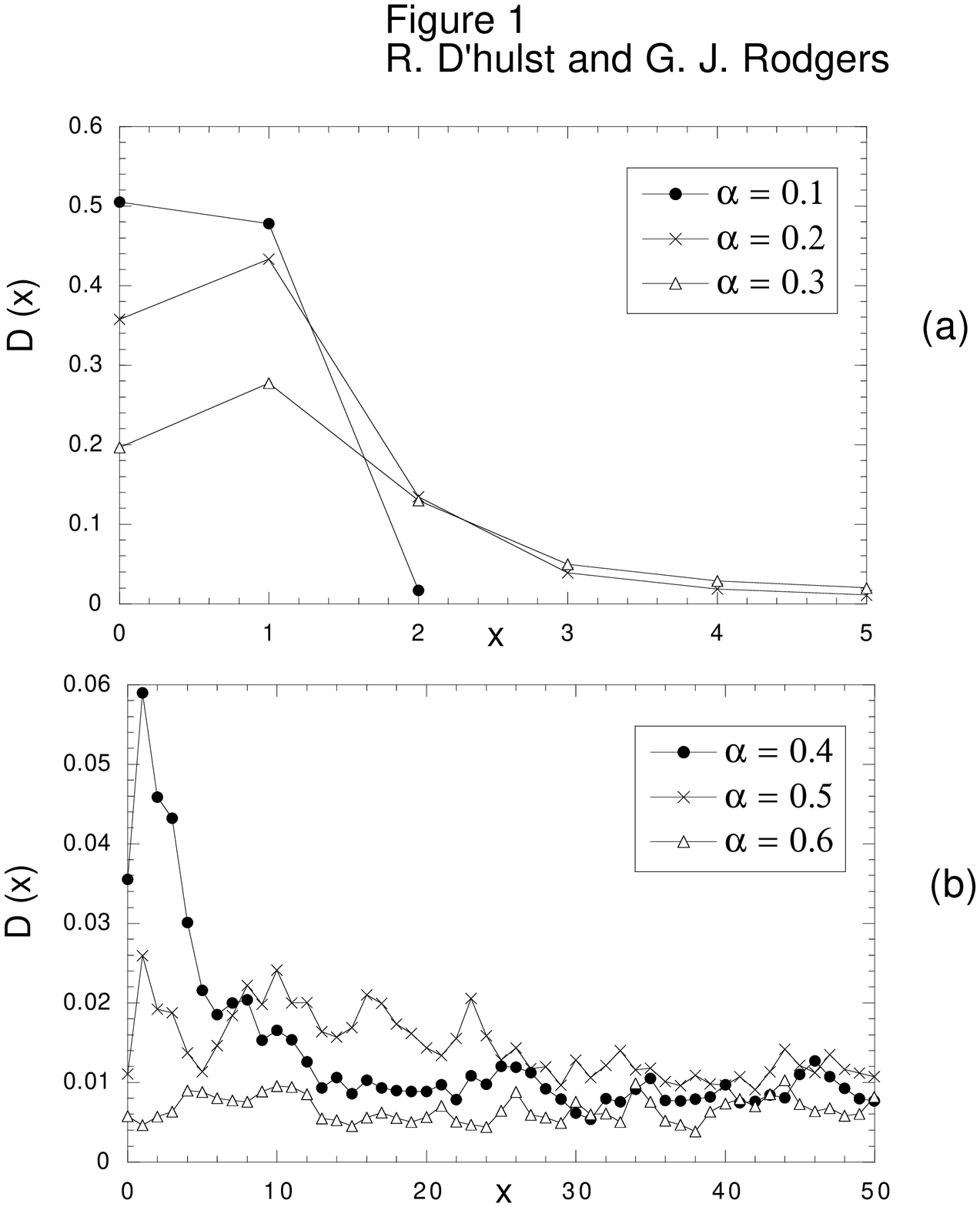}
\end{picture}
\newpage

\begin{picture}(21,27)(3,-4)
\epsfig{file=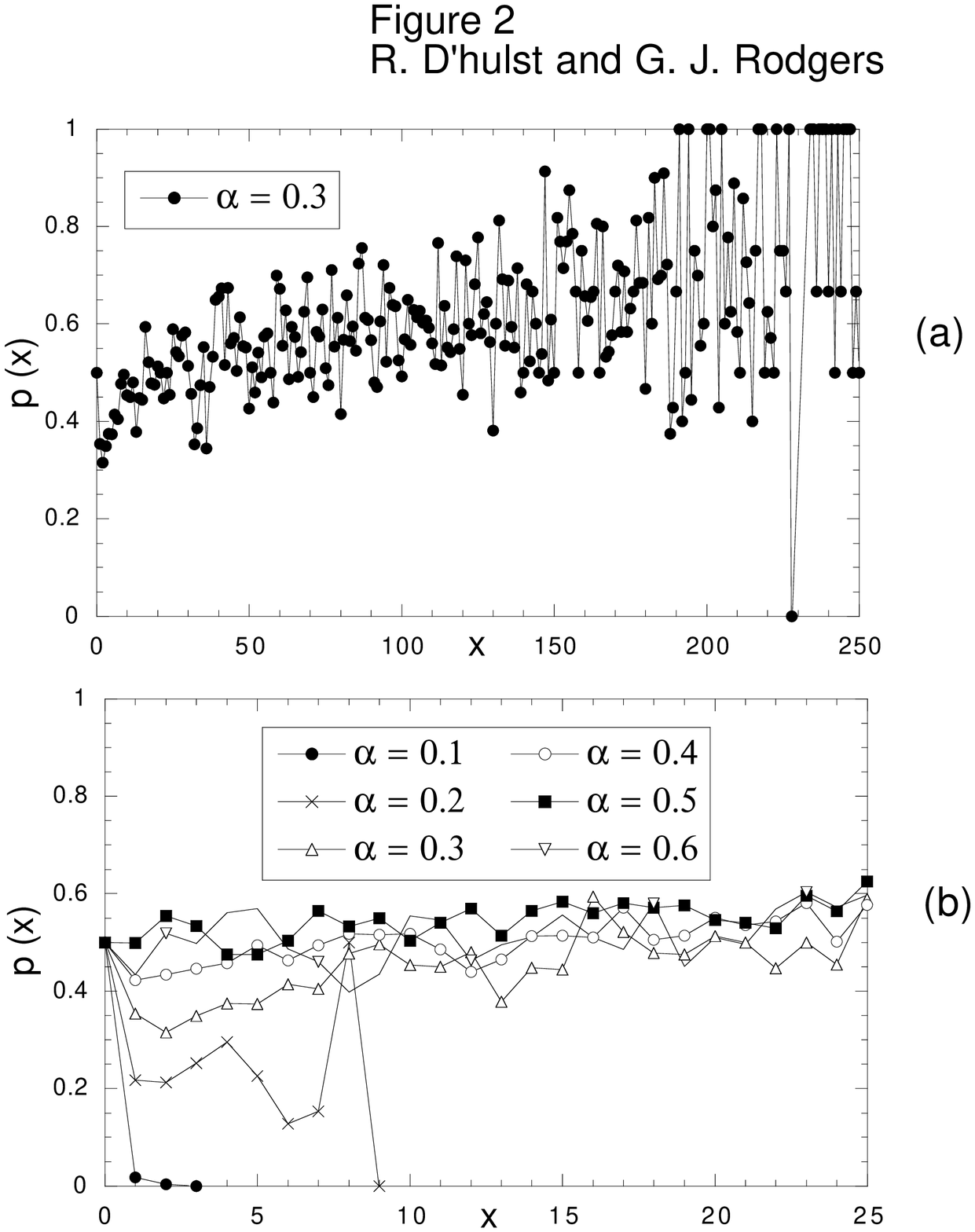}
\end{picture}
\newpage

\begin{picture}(21,27)(3,-7)
\epsfig{file=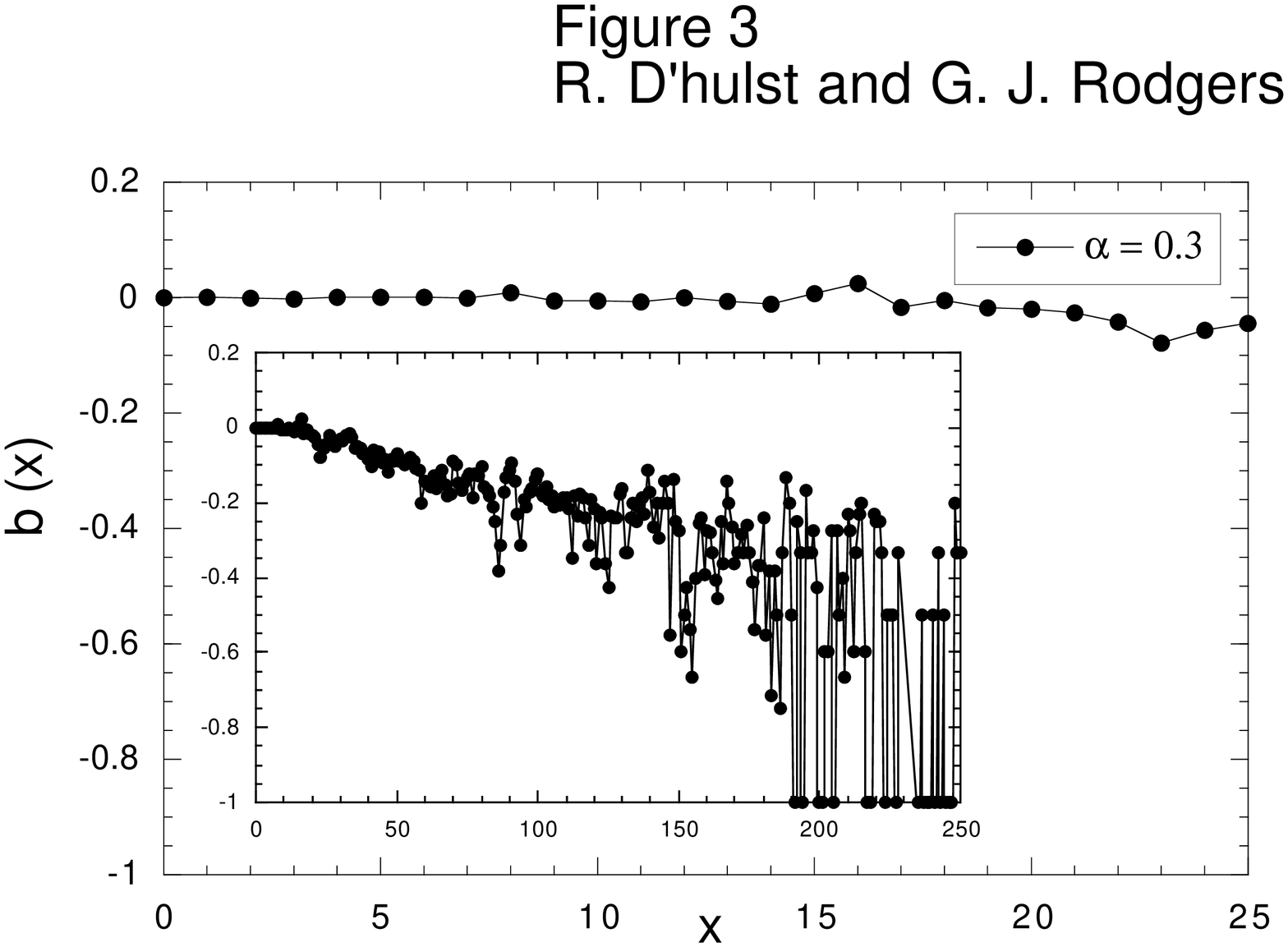,angle=90}
\end{picture}
\newpage

\begin{picture}(21,27)(3,-7)
\epsfig{file=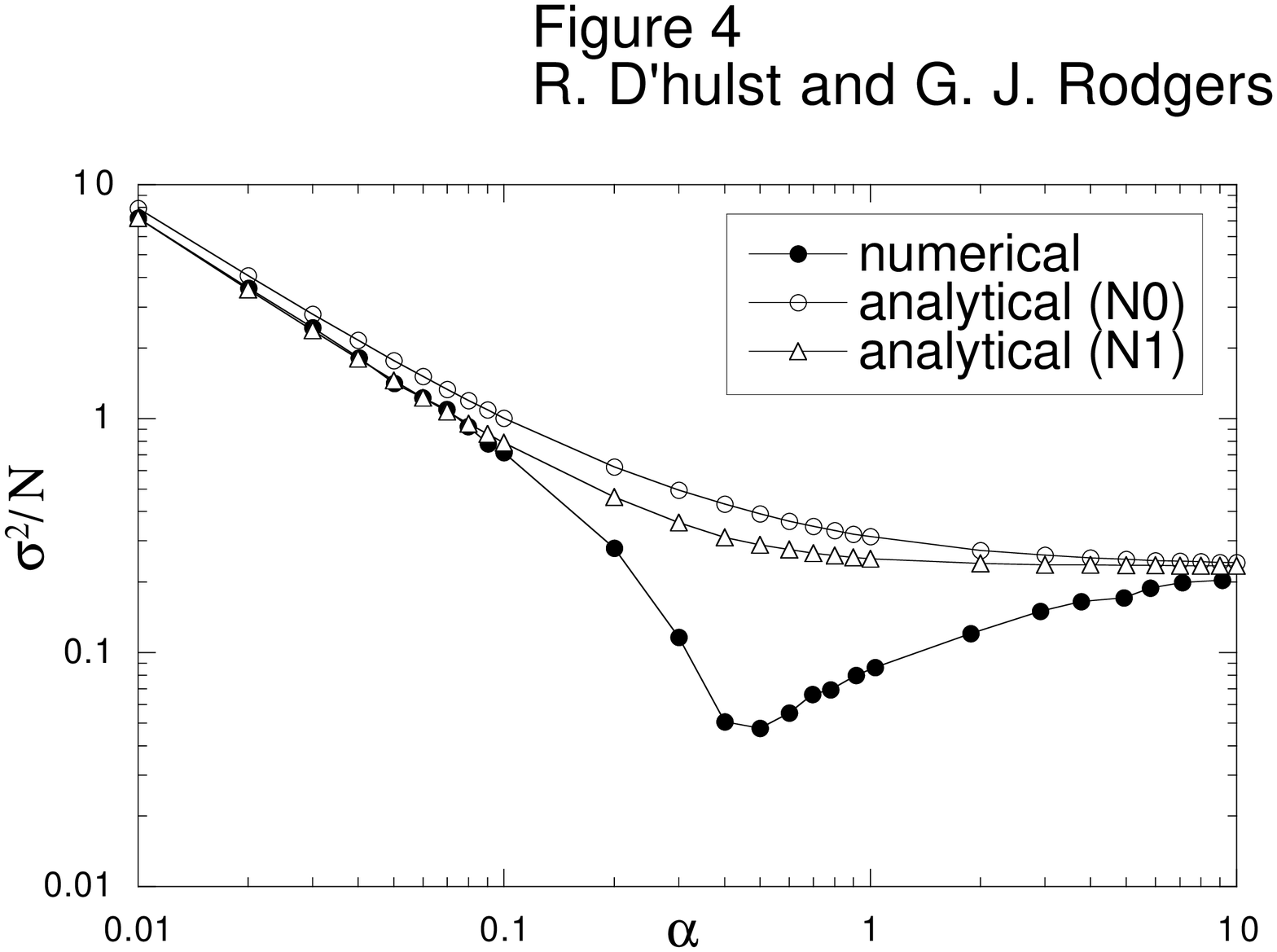,angle=90}
\end{picture}


\newpage
\begin{thebibliography}{99}
\bibitem{challet9798}D. Challet and Y.-C. Zhang, Physica A {\bf 246}, (1997) 407 (adap-org/9708006).
\bibitem{zhang98}Y.-C. Zhang, Europhysics News {\bf 29}, (1998) 51 (cond-mat/9803308).
\bibitem{tous}N. F. Johnson, P. M. Hui, R. Jonson and T. S. Lo, Phys. Rev. Lett. {\bf 82}, (1999) 3360 (cond-mat/9810142); A. Cavagna, J. Garrahan, I. Giardina and D. Sherrington, cond-mat/9907296; R. D'Hulst and G. J. Rodgers, adap-org/9904003.
\bibitem{savit97}R. Savit, R. Manuca and R. Riolo, Phys. Rev. Lett. {\bf 82}, (1999) 2203 (adap-org/9712006).
\bibitem{dhulst99}R. D'Hulst and G. J. Rodgers, Physica A {\bf 270}, (1999) 514 (adap-org/9902001).
\bibitem{challet99-2}D. Challet, M. Marsili and R. Zecchina, cond-mat/9904392.
\bibitem{johnson98-1}N. F. Johnson, M. Hart and P. M. Hui, Physica A {\bf 269}, (1999) 1 (cond-mat/9811227).
\end{thebibliography}
\end{document}